# EDOT polymerization at photolithographically patterned functionalized graphene

*Petr Kovaříček,[1] Karolina Drogowska,[1] Zuzana Komínková Melníková,[1,2] Václav Blechta,[1,2] Zdeněk Bastl,[1] Daniel Gromadzki,[1] Michaela Fridrichová,[1] Martin Kalbáč[1,*]*

[1] J. Heyrovsky Institute of Physical Chemistry of the Academy of Sciences of the Czech Republic, Dolejškova 2155/3, 182 23 Praha, Czech Republic,
[2] Palacký University, Department of Physical Chemistry, Faculty of Science, 17.listopadu 12, 771 46 Olomouc, Czech Republic

**Abstract**

Graphene possesses unique features that make it attractive for nanotechnology. Functional devices often require combination of several materials with specific functions, and graphene-polymer composites are one of them. Herein, we report on the preparation of PEDOT:Graphene bilayers by in situ polymerization of EDOT on covalently functionalized graphene. The polymerization proceeds exclusively on the grafted graphene, and patterned structures with high spatial resolution down to 3 μm could have been prepared. The composite exhibits enhanced efficiency of electrochemical doping compared to pristine graphene, unsymmetrical transport characteristic with very good hole-transporting properties and efficient electronic communication between the two materials.

**1. Introduction**

Graphene, due to its outstanding properties, shows great promise for a wide variety of applications in nanotechnology.[1–5] With its unique, semi-metallic band structure,[6,7] graphene opens the gate to modern and prospective purposes because its electronic nature can be easily manipulated by fabrication of various heterogenenous structures[4,5,8] with manifold functional materials, such as, e.g., conductive polymers and poly(3,4-ethylenedioxythiophene) (PEDOT) as their epitomical example.[9–15] Successful advancement in downsizing graphene-based functional devices requires high spatial resolution of the patterned structures,[16,17] which faces substantial challenges when precise composite devices are fabricated.[18–20] PEDOT layers are routinely prepared by spincoating the commercial blend with polystyrenesulfonate (PSS),[9–12] but such a process does not allow for exclusive film formation at desired areas. Procedures employing electrochemical polymerization of PEDOT at various (carbon-based) electrodes have been demonstrated as well,[13,21] but this approach requires tedious and very precise electrical contacting of the material, which limits spatial resolution. Instead, here we demonstrate selective formation of PEDOT films at functionalized monolayer graphene, which is photolithographically patterned prior to the polymerization reaction. This process affords superb spatial resolution and high selectivity of the film formation as well as unlimited variability and thus is considered highly perspective when designing 2D structures. The presented protocol is of particular interest for fabrication of various (opto-)electronic devices in which the electron-blocking/hole-transporting properties of PEDOT and very high conductivity of graphene are critically required, such as in the fields of photovoltaics, LEDs or LETs fabrication, or sensors.

---

[*] Tel.: +420-266053804, E-mail: martin.kalbac@jh-inst.cas.cz

## 4. Experimental Section

Fitting of Raman spectra was performed by own-built Matlab® script using the symmetrical pseudo-Voigt signal shape, and strain-doping analysis was performed according to the literature procedure.[22]

**Graphene growth and transfer**

Graphene was grown on copper by CVD and transferred to $SiO_2$/Si substrates, as described previously.[23,24] In brief, a polycrystalline copper foil was annealed for 20 min at 1000 °C in hydrogen. Graphene was then grown from 1 sccm $CH_4$ for 35 min and then annealed for another five minutes in $H_2$ atmosphere. The sample was then cooled to room temperature and transferred to a Si/$SiO_2$ substrate using the nitrocellulose-based technique.[25]

**Photolithography – graphene patterns**

Graphene on substrate was covered with photoresist AZ 6632 by spincoating, aligned with the mask, exposed to the mercury short arc lamp light and developed by the AZ726MIF. Then graphene was etched by oxygen plasma and the polymer mask was removed by the NI555 stripper. Samples were then washed with *i*-propanol and dried in a stream of nitrogen.

**Graphene functionalization**

In a 50 mL beaker, 75 mg of sulfanilic acid was suspended in 5 mL of deionized $H_2O$ (> 18 $M\Omega$ $cm^{-1}$) and a solution of 30 mg of $NaNO_2$ in 5 mL of deionized water was added upon vigorous stirring at room temperature. When all sulfanilic acid dissolved, graphene on substrate was immersed in the reaction mixture and kept at room temperature for 30 minutes. Then the wafer was removed from the solution, neutralized by soaking into 2 % aqueous HCl for 2 minutes and washed repeatedly (3x) by 5 mL of deionized water and 5 mL of UV-VIS grade methanol. Additional analytical data is provided in the SI and labeled as **FGr1**.

**EDOT polymerization**

In a 50 mL beaker 60 mg of $FeCl_3 \cdot 6\ H_2O$ was diluted either in 25 mL of acetonitrile or 10 mL of propylene carbonate and 32 mg of 3,4-ethylenedioxythiophene was added via syringe while stirring. In the case of acetonitrile solutions, deep blue color evolved immediately. Functionalized graphene on substrate was then immersed in the solution for a given time (20 – 80 minutes) at room temperature, then removed and washed thoroughly with UV-VIS grade methanol.

## 2. Results and discussion

Single layer graphene was grown by the chemical vapor deposition method on copper and transferred onto a silicon substrate with 300 nm thick layer of $SiO_2$ ($Si/SiO_2$) as described earlier.[23–27] This graphene sheet shows the typical signature in Raman spectra,[28–31] namely, the G and 2D mode at 1595 and 2680 $cm^{-1}$, respectively, without observable D band, indicating very low amount of defects in the transferred monolayer. In the first step, graphene was covalently functionalized by 4-phenylsulfonate using the standard procedure via diazonium salt of sulfanilic acid (**Figure 1a**, details in SI).[31–34] Successful grafting is demonstrated by the emerging D and D' modes at 1345 and 1621 cm$^{-1}$, respectively, while the intensity of the 2D mode decreases, which is in agreement with the reported higher reactivity of single layer graphene as compared to multilayers.[35] Uniformity of the functionalization was verified by Raman mapping over a 20 x 20 μm and plotting the relative D/G band intensity (see SI) from which also the extent of functionalization can be determined according to the published model.[36–38] In our case, the median of distance between functional groups ($L_D$ value) was found at 2.38 nm corresponding to defect density $\sigma = 1.77 \times 10^{13}$ cm$^{-2}$.

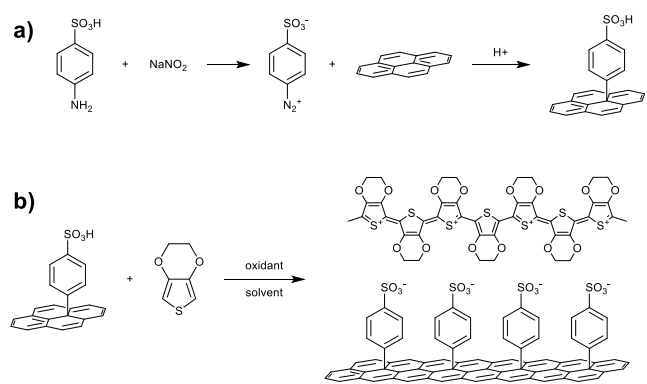

**Figure 1.** a) Scheme of graphene functionalization using the diazonium protocol. Zwitterionic diazonium is formed upon mixing sulfanilic acid and sodium nitrite in water, as indicated by dissolution of the sparsely soluble acid and development of yellow-brown color. Graphene immersed into the solution reacts within 30 minutes to provide the functionalized material bearing sulfonate groups. b) Scheme of EDOT polymerization at functionalized graphene. Oxidative α-α'coupling of thiophene rings by the action of iron(III) chloride affords PEDOT bearing some positive charge, which is compensated by negative sulfonate groups on graphene.

PEDOT is typically commercialized as a blend with PSS, which helps both to stabilize and solubilize the polymer. In our approach, we have employed the phenylsulfonate grafted graphene as the stabilizing matrix, which should in turn selectively promote polymerization of PEDOT at functionalized graphene compared with the bare substrate surface (**Figure 1b**). We have investigated the polymerization reaction under various conditions, and we have evaluated the thickness of the PEDOT film together with the selectivity of the polymerization at the functionalized graphene in contrast with the bare Si/SiO$_2$ substrate for each of these conditions by AFM and Raman spectroscopy imaging.

First, we attempted the polymerization of 3,4-ethylenedioxythiophene (EDOT) in water as a solvent, which is used for the PEDOT:PSS blends solutions.[39] The successful formation of the polymer film was demonstrated by the appearance of the characteristic PEDOT Raman signals at 1433, 1513, 1565, 2862 cm$^{-1}$; however, due to low solubility of EDOT in water, large non-homogeneities were observed on the surface. To overcome this issue, we have employed two organic solvents, namely, acetonitrile and propylene carbonate, in which the polymerization of EDOT has been previously demonstrated.[21,40,41] Indeed, the reactant and the oxidant (FeCl$_3$ · 6 H$_2$O) are miscible/soluble in these solvents, so solutions of up to 1.0 M concentration could have been prepared, but for our experiments, we have used 9 and 22 mM concentration of both components in acetonitrile and propylene carbonate, respectively. A large difference in the polymerization kinetics between the two solvents has been observed: Whereas the reaction mixture in acetonitrile turned deep blue instantaneously upon mixing, in propylene carbonate the color was slowly developing, starting from yellow and turning toward greenish and then blue over an hour.

We have also varied the reaction time in both solvents from 20 to 80 minutes and determined the thickness of the polymer film on graphene by AFM profiling. It has been found that under the conditions used, the formation of PEDOT occurred exclusively on the functionalized graphene (**Figure 2**). The choice of solvent translated into different polymer film thickness: In acetonitrile, the composite bilayer PEDOT:graphene thickness falls between 4-6 nm, and in propylene carbonate it was estimated to 2-3 nm (**Figure 3**). On the other hand, no trends in thickness change as a function of reaction time have been observed, probably due to the small difference and limited precision of the AFM profiling. X-ray photoelectron spectra of the prepared PEDOT:graphene composite bilayers were recorded and gave an elemental ratio C:S 95:5 respectively. Sulfur atoms were found to exist in two chemical states characterized by the S 2p3/2 binding energies of 163.4 and 166.6 eV corresponding to the formal 2⁻ states in the thiophene moiety and 6⁺ state in the phenylsulfonyl group, respectively, which is in agreement with previous reports.[42] Trace amounts of the reaction mixture were also detected (<1 %), namely iron and chlorine arising from the oxidant, and nitrogen from the acetonitrile solvent.

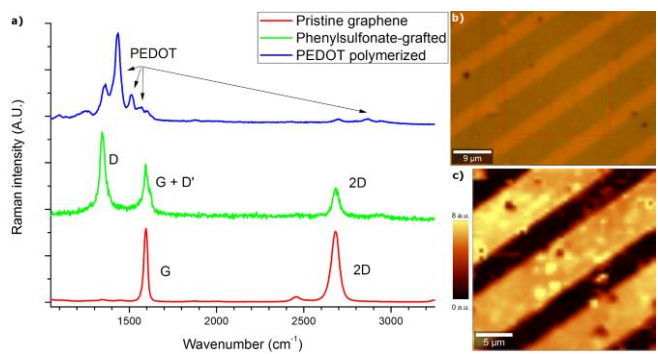

**Figure 2.** Raman spectroscopy of PEDOT polymerized at functionalized graphene measured with 532 nm excitation laser. a) Spectra measured after each step of the procedure (from bottom to top): pristine graphene after transfer onto Si/SiO$_2$ wafer showing the characteristic G and 2D bands with relative amplitudes and full widths at half maxima typical for graphene deposited on SiO$_2$ substrate (red trace). Grafting of phenylsulfonate groups leads to appearance of the D and D' modes (green trace), and the successful polymerization of EDOT is demonstrated by its characteristic signals at 1433, 1513, 1565, 2862 cm$^{-1}$ (blue trace). b) Optical image of the photolithographically patterned graphene, functionalized with phenylsulfonate and polymerized PEDOT film. Area subjected to Raman mapping is indicated by the red square. c) Raman map showing the intensity of the PEDOT dominant signal at around 1433 cm$^{-1}$. Graphene stripes are 5 μm wide and separated by a 3 μm gap between them.

The exclusive formation of PEDOT at functionalized graphene compared with bare substrate tempted us to probe the limits of the polymer spatial distribution prepared by this method. In this vein, we have prepared graphene stripes by photolithography using positive photoresist and oxygen plasma etching of graphene (see SI for details). The dimensions of stripes were given by the mask and ranged from 3 to 10 μm with 3 to 10 μm separation between them. These structures were subjected to the sequence of phenylsulfonate functionalization and EDOT polymerization in acetonitrile described above and analyzed by Raman mapping (**Figure 2c**). The PEDOT signal (1433 cm$^{-1}$) is only present on graphene stripes, and the 3 μm wide gap prevents formation of a continuous polymer layer, which thus shows high selectivity of the PEDOT:graphene heterostructure formation with superb spatial resolution.

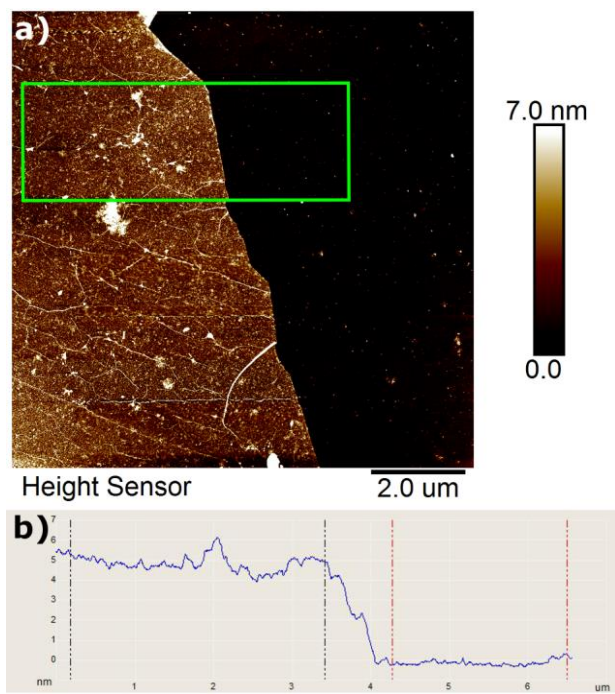

**Figure 3.** AFM image (a) and height profile (b) of the PEDOT:Graphene composite prepared by covalent functionalization of graphene and selective polymer film formation on it. The height of the composite sheet was calculated from the area highlighted in green, showing an average thickness of the PEDOT:Graphene bilayer of about 5 nm.

We have performed Raman mapping of all the prepared samples over areas up to 45 x 35 μm large. By fitting of the indicative signals in Raman maps, we were able to evaluate important parameters such as homogeneity of the PEDOT film and strain-doping analysis of graphene. It has been found that graphene is slightly but homogeneously strained (< 0.05 %) and the polymer layer is nicely uniform as shown by the relative signal intensity (**Figure 2**). The quality of the film was further verified by scanning electron microscopy (SEM, see SI). It can also be seen that the PEDOT Raman signal intensity correlates with the level of graphene doping (around 1-4 x $10^{13}$ $cm^{-2}$), indicating the interaction between the two oppositely charged layers (more details can be found in the SI).

The PEDOT:graphene composite bilayer was also investigated by in situ Raman spectroelectrochemistry.[43] This method allows study of changes in the electronic structure of the material caused by doping when the graphene doping is directly controlled by applied electrode potential. Unlike chemical doping, this method enables easy, reversible and precisely tuned doping of graphene in the range of -1.5 V and 1.5 V (step 0.3 V). **Figure 4** shows the development of the potential-dependent Raman spectra for the single layer graphene covered with PEDOT layer on Si/SiO$_2$ substrate (see also SI for cyclic voltammograms and optical images).

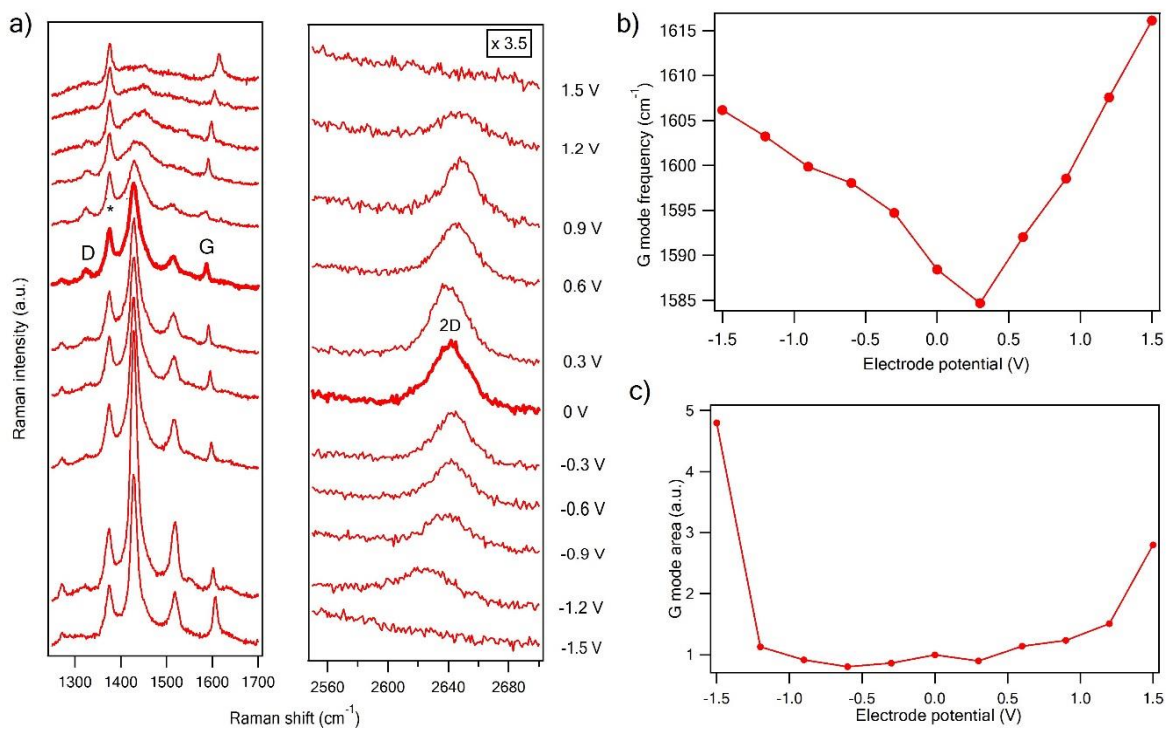

**Figure 4.** In-situ Raman spectroelectrochemical data of the functionalized single layer graphene with the PEDOT film grown on top on a Si/SiO$_2$ substrate with focusing on the D, G and 2D modes measured in the potential range between -1.5 V and 1.5 V. The D, G and 2D modes' peaks are marked in the Raman spectrum. The bold line denotes the Raman spectrum at electrode potential 0 V. The Raman spectra are excited by 1.96 eV (633 nm) laser radiation. The peak at 1375 cm$^{-1}$ related to the electrolyte is marked by an asterisk, and it is not affected by doping. Dependence of b) the G mode position shift and c) the G mode area (intensity) on applied electrode potential. The electrode potentials are given versus the Ag/Ag$^+$ reference electrode.

The first panel (1250-1700 cm$^{-1}$) of **Figure 4a** shows the development of the PEDOT peaks (1270, 1428 and 1515 cm$^{-1}$) and the graphene D (1323 cm$^{-1}$) and G (1588 cm$^{-1}$) modes at positive and negative doping. The band at 1270 cm$^{-1}$ corresponds to vibration of the Cα−Cα bond and the most intense bands at 1428 cm$^{-1}$ and 1515 cm$^{-1}$ to symmetric and asymmetric vibrations of Cα=Cβ bonding, respectively. It has been reported that the intensity of the PEDOT Raman bands changes significantly with applied electrode potentials.[44] The behavior of PEDOT peaks in the sample of single layer graphene covered by PEDOT layer agrees with the previously published data,[39,44] i.e. their intensity increases steadily with increasing negative potential from 0 V to -1.2 V and decreases slightly beyond. Conversely, the PEDOT peaks' intensity drops down monotonically with applied positive doping and almost vanishes at the highest positive potential (1.5 V), while Raman shift of all peaks remains unchanged during both positive and negative doping. Interestingly, the intensity of the D mode is decreased more at negative than at positive doping, which could be due to the interaction with PEDOT layer. While for positive doping the polymer becomes transparent, for negative doping, its absorbance increases. Therefore, the PEDOT layer can also prevent emission of scattered light from graphene, and the signal from the graphene D mode is thus decreased. The second panel of **Figure 4a** shows the potential-dependence development of the 2D mode, which decreases monotonically with increasing electrode potential in both cathodic and anodic directions. The shifts of 2D mode are influenced by changes in the C-C bond strength, the electron-phonon coupling and electron-electron interactions. As a result, positive doping increases and negative decreases the 2D mode Raman shift, which was also observed in our case, agreeing with previously reported data.[45]

The detail development of the G mode during electrochemical doping is shown in **Figure 4b** and **4c**. The trends of the G mode Raman shift (**Figure 4b**) agree with previously published results.[45] The G mode is upshifted at both positive and negative doping. This G mode shift change relates to the change in the C-C bond strength[46] and the renormalization of phonon energy.[47] The G mode position at 0 V is 1588 cm$^{-1}$ and increases monotonically with the increasing electrode potential, finally reaching shifts of 28 cm$^{-1}$ at 1.5 V and of 18 cm$^{-1}$ at -1.5 V electrode potential. **Figure 4c** shows the dependence of the G mode area (intensity) on applied electrode potential. As realized in earlier instances,[48,49] the G mode area is increased at both high positive and negative electrode potentials that are assigned to quantum interference between Raman pathways in graphene.[50] In our data, the G mode area is increased about 3 times at electrode potential 1.5 V compared with the G mode area at 0 V, and about 5 times at -1.5 V. Typically, a higher increase in the G mode area is observed at the highest positive electrode potential (1.5 V) because the charge injected by application of negative potentials is partially compensated for by the substrate. However, in this case, the PEDOT functionalization of graphene enhances the efficiency of graphene negative doping, which in turn blocks electronic transitions and ultimately leads to higher G mode area at -1.5 V electrode potential (**Figure 4c**).

The combination of graphene electrical conductivity and the hole-transporting/injection features of PEDOT is intensely studied for a wide range of (opto-)electronic applications, which turned us to investigation of the actual electrical performance of our composite system. For this purpose we have prepared graphene field-effect transistor (GFET, details in the SI), functionalized the material with phenylsulfonyl groups and *in situ* polymerized PEDOT on top. Electrical parameters were determined for both pristine and functionalized samples at elevated temperature in inert atmosphere to effectively remove moisture and other possible adsorbates, which were significantly increasing the experimental noise of the measurement and shifting the Dirac point beyond accessible region.

The GFET setup of the electrical measurements allows to determine crucial parameters such as doping level and charge carrier mobility from the shape of the I-V curve. Pristine (not annealed) graphene on Si/SiO$_2$ wafer is p-doped by the substrate and exhibit Dirac point at 22 V resulting in mobility of 119 vs. 27 cm$^2$ V$^{-1}$ s$^{-1}$ for holes and electrons, respectively. Introduction of sulfonylphenyl groups via the diazonium leads to drastic decrease of hole mobility to 0.6 cm$^2$ V$^{-1}$ s$^{-1}$ and shift of the Dirac point beyond accessible gate voltage region. The transport characteristic after EDOT polymerization are presented in **Figure 5**. The asymmetric shape of these curves clearly indicates inequivalent mobility for holes and electrons. These were extracted from the linear parts of the transport characteristics[51] amounting to 1010 cm$^2$ V$^{-1}$ s$^{-1}$ and 80 cm$^2$ V$^{-1}$ s$^{-1}$ for holes and electrons respectively. From the I-V plot also the position of the Dirac point can be calculated. When measured under ambient conditions (r.t., air) it appeared above 100 V, however at 50 °C in nitrogen its position was found at 60 V. This value is consistent with significant p-doping as expected for a positively charge polymer.[52] Interestingly, although the Dirac point is strongly shifted, the hole mobility is very similar under both experimental conditions. To evaluate temporal stability of the composite material, extended measurement of the transport characteristics was conducted for 24 hours and no significant change in the properties was found. Altogether these observations show that the ambipolar charge carrier feature of graphene can be easily tuned to give strongly asymmetric charge carrier mobilities of the resulting composite material by more than an order of magnitude.

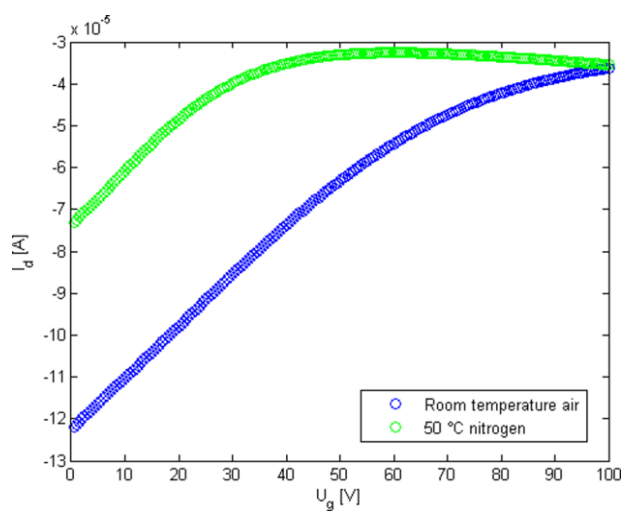

**Figure 5.** Measurement of PEDOT:Graphene GFET device under ambient conditions (r.t., air, blue) and at 50 °C in nitrogen (green). Dirac point can be seen at around 60 V at elevated temperature under N$_2$, but ambient atmosphere shifts it out of the accessible range. Much higher mobility of holes (1010 cm$^2$ V$^{-1}$ s$^{-1}$) vs. mobility of electrons (80 cm$^2$ V$^{-1}$ s$^{-1}$) can be extracted from the linear parts of the asymmetric transport characteristics.

## 3. Conclusion

In summary, we have shown the process of EDOT polymerization at functionalized graphene patterns prepared by photolithography. Graphene, grown by CVD method on copper and transferred on Si/SiO$_2$ wafer, was grafted with phenylsulfanyl groups that promote the reaction and stabilize its product by complexation. The reaction proceeds exclusively on the functionalized graphene, allowing formation of functional patterns via photolithography and graphene etching. The spatial resolution of the polymer distribution is limited by the precision of the photolithography process, which in our case provided structures with dimensions as small as 3 μm. The obtained PEDOT:graphene composite was studied by in situ Raman spectroelectrochemistry, revealing strong interactions in the bilayer and enhanced efficiency of negative doping of the material. The material was also investigated for its electronic properties revealing very good electron-blocking/hole-transporting properties and shifting of the Dirac point in response to temperature and atmospheric composition. We are convinced that the presented protocol will significantly help in preparation of various (opto-)electronic devices in which asymmetric charge carrier mobilities are required, such as photovoltaics or LED fabrication.

**Supporting Information**
Synthetic protocols, sample preparation procedures and additional analytical data are provided in the Supplementary information file.

**Acknowledgements**
The work was supported by ERC-CZ project No. LL1301. P.K. thanks the ASCR PPPLZ program for funding.

**References**
[1] A.K. Geim, K.S. Novoselov, The rise of graphene, Nat. Mater. 6 (2007) 183–191. doi:10.1038/nmat1849.
[2] K.S. Novoselov, A.K. Geim, S.V. Morozov, D. Jiang, Y. Zhang, S.V. Dubonos, I.V. Grigorieva, A.A. Firsov, Electric Field Effect in Atomically Thin Carbon Films, Science. 306 (2004) 666–669. doi:10.1126/science.1102896.
[3] K.S. Novoselov, A.K. Geim, S.V. Morozov, D. Jiang, M.I. Katsnelson, I.V. Grigorieva, S.V. Dubonos, A.A. Firsov, Two-dimensional gas of massless Dirac fermions in graphene, Nature. 438 (2005) 197–200. doi:10.1038/nature04233.
[4] T. Ramanathan, A.A. Abdala, S. Stankovich, D.A. Dikin, M. Herrera-Alonso, R.D. Piner, D.H. Adamson, H.C. Schniepp, X. Chen, R.S. Ruoff, S.T. Nguyen, I.A. Aksay, R.K. Prud'Homme, L.C. Brinson, Functionalized graphene sheets for polymer nanocomposites, Nat. Nanotechnol. 3 (2008) 327–331. doi:10.1038/nnano.2008.96.
[5] V. Dhand, K.Y. Rhee, H. Ju Kim, D. Ho Jung, V. Dhand, K.Y. Rhee, H. Ju Kim, D. Ho Jung, A Comprehensive Review of Graphene Nanocomposites: Research Status and Trends, J. Nanomater. 2013 (2013) e763953. doi:10.1155/2013/763953.
[6] M.F. Craciun, S. Russo, M. Yamamoto, J.B. Oostinga, A.F. Morpurgo, S. Tarucha, Trilayer graphene is a semimetal with a gate-tunable band overlap, Nat. Nanotechnol. 4 (2009) 383–388. doi:10.1038/nnano.2009.89.

[7]  G. Lu, K. Yu, Z. Wen, J. Chen, Semiconducting graphene: converting graphene from semimetal to semiconductor, Nanoscale. 5 (2013) 1353–1368. doi:10.1039/c2nr32453a.

[8]  J.R. Potts, D.R. Dreyer, C.W. Bielawski, R.S. Ruoff, Graphene-based polymer nanocomposites, Polymer. 52 (2011) 5–25. doi:10.1016/j.polymer.2010.11.042.

[9]  W. Hong, Y. Xu, G. Lu, C. Li, G. Shi, Transparent graphene/PEDOT–PSS composite films as counter electrodes of dye-sensitized solar cells, Electrochem. Commun. 10 (2008) 1555–1558. doi:10.1016/j.elecom.2008.08.007.

[10] M. Zhang, W. Yuan, B. Yao, C. Li, G. Shi, Solution-Processed PEDOT:PSS/Graphene Composites as the Electrocatalyst for Oxygen Reduction Reaction, ACS Appl. Mater. Interfaces. 6 (2014) 3587–3593. doi:10.1021/am405771y.

[11] D. Yoo, J. Kim, J.H. Kim, Direct synthesis of highly conductive poly(3,4-ethylenedioxythiophene):poly(4-styrenesulfonate) (PEDOT:PSS)/graphene composites and their applications in energy harvesting systems, Nano Res. 7 (2014) 717–730. doi:10.1007/s12274-014-0433-z.

[12] Y. Liu, B. Weng, J.M. Razal, Q. Xu, C. Zhao, Y. Hou, S. Seyedin, R. Jalili, G.G. Wallace, J. Chen, High-Performance Flexible All-Solid-State Supercapacitor from Large Free-Standing Graphene-PEDOT/PSS Films, Sci. Rep. 5 (2015) 17045. doi:10.1038/srep17045.

[13] C.-Y. Chu, J.-T. Tsai, C.-L. Sun, Synthesis of PEDOT-modified graphene composite materials as flexible electrodes for energy storage and conversion applications, Int. J. Hydrog. Energy. 37 (2012) 13880–13886. doi:10.1016/j.ijhydene.2012.05.017.

[14] Z. Liu, K. Parvez, R. Li, R. Dong, X. Feng, K. Müllen, Transparent Conductive Electrodes from Graphene/PEDOT:PSS Hybrid Inks for Ultrathin Organic Photodetectors, Adv. Mater. 27 (2015) 669–675. doi:10.1002/adma.201403826.

[15] G. Sonmez, Polymeric electrochromics, Chem. Commun. (2005) 5251–5259. doi:10.1039/B510230H.

[16] G. Fiori, F. Bonaccorso, G. Iannaccone, T. Palacios, D. Neumaier, A. Seabaugh, S.K. Banerjee, L. Colombo, Electronics based on two-dimensional materials, Nat. Nanotechnol. 9 (2014) 768–779. doi:10.1038/nnano.2014.207.

[17] K. Kostarelos, K.S. Novoselov, Graphene devices for life, Nat. Nanotechnol. 9 (2014) 744–745. doi:10.1038/nnano.2014.224.

[18] H. Yoon, J. Jang, Conducting-Polymer Nanomaterials for High-Performance Sensor Applications: Issues and Challenges, Adv. Funct. Mater. 19 (2009) 1567–1576. doi:10.1002/adfm.200801141.

[19] B. Charlot, G. Sassine, A. Garraud, B. Sorli, A. Giani, P. Combette, Micropatterning PEDOT:PSS layers, Microsyst. Technol. 19 (2012) 895–903. doi:10.1007/s00542-012-1696-5.

[20] B.H. Lee, J.-H. Lee, Y.H. Kahng, N. Kim, Y.J. Kim, J. Lee, T. Lee, K. Lee, Graphene-Conducting Polymer Hybrid Transparent Electrodes for Efficient Organic Optoelectronic Devices, Adv. Funct. Mater. 24 (2014) 1847–1856. doi:10.1002/adfm.201302928.

[21] K. Krukiewicz, J.S. Bulmer, D. Janas, K.K.K. Koziol, J.K. Zak, Poly(3,4-ethylenedioxythiophene) growth on the surface of horizontally aligned MWCNT electrode, Appl. Surf. Sci. 335 (2015) 130–136. doi:10.1016/j.apsusc.2015.02.039.

[22] J.E. Lee, G. Ahn, J. Shim, Y.S. Lee, S. Ryu, Optical separation of mechanical strain from charge doping in graphene, Nat. Commun. 3 (2012) 1024. doi:10.1038/ncomms2022.

[23] A. Reina, X. Jia, J. Ho, D. Nezich, H. Son, V. Bulovic, M.S. Dresselhaus, J. Kong, Large Area, Few-Layer Graphene Films on Arbitrary Substrates by Chemical Vapor Deposition, Nano Lett. 9 (2009) 30–35. doi:10.1021/nl801827v.

[24] J. Ek-Weis, S. Costa, O. Frank, M. Kalbac, Heating Isotopically Labeled Bernal Stacked Graphene: A Raman Spectroscopy Study, J. Phys. Chem. Lett. 5 (2014) 549–554. doi:10.1021/jz402681n.

[25] T. Hallam, N.C. Berner, C. Yim, G.S. Duesberg, Strain, Bubbles, Dirt, and Folds: A Study of Graphene Polymer-Assisted Transfer, Adv. Mater. Interfaces. 1 (2014) 1400115. doi:10.1002/admi.201400115.

[26] J. Ek Weis, S.D. Costa, O. Frank, Z. Bastl, M. Kalbac, Fluorination of Isotopically Labeled Turbostratic and Bernal Stacked Bilayer Graphene, Chem. – Eur. J. 21 (2015) 1081–1087. doi:10.1002/chem.201404813.

[27] P. Kovaříček, Z. Bastl, V. Valeš, M. Kalbac, Covalent Reactions on Chemical Vapor Deposition Grown Graphene Studied by Surface-Enhanced Raman Spectroscopy, Chem. – Eur. J. 22 (2016) 5404–5408. doi:10.1002/chem.201504689.

[28] A.C. Ferrari, J.C. Meyer, V. Scardaci, C. Casiraghi, M. Lazzeri, F. Mauri, S. Piscanec, D. Jiang, K.S. Novoselov, S. Roth, A.K. Geim, Raman Spectrum of Graphene and Graphene Layers, Phys. Rev. Lett. 97 (2006) 187401. doi:10.1103/PhysRevLett.97.187401.

[29] L.M. Malard, M.A. Pimenta, G. Dresselhaus, M.S. Dresselhaus, Raman spectroscopy in graphene, Phys. Rep. 473 (2009) 51–87. doi:10.1016/j.physrep.2009.02.003.

[30] A. Jorio, M.S. Dresselhaus, R. Saito, G. Dresselhaus, Raman spectroscopy in graphene related systems, Wiley-VCH, Weinheim, 2011.

[31] G.L.C. Paulus, Q.H. Wang, M.S. Strano, Covalent Electron Transfer Chemistry of Graphene with Diazonium Salts, Acc. Chem. Res. 46 (2013) 160–170. doi:10.1021/ar300119z.

[32] J.R. Lomeda, C.D. Doyle, D.V. Kosynkin, W.-F. Hwang, J.M. Tour, Diazonium Functionalization of Surfactant-Wrapped Chemically Converted Graphene Sheets, J. Am. Chem. Soc. 130 (2008) 16201–16206. doi:10.1021/ja806499w.

[33] P. Huang, L. Jing, H. Zhu, X. Gao, Diazonium Functionalized Graphene: Microstructure, Electric, and Magnetic Properties, Acc. Chem. Res. 46 (2013) 43–52. doi:10.1021/ar300070a.

[34] D. Bouša, O. Jankovský, D. Sedmidubský, J. Luxa, J. Šturala, M. Pumera, Z. Sofer, Mesomeric Effects of Graphene Modified with Diazonium Salts: Substituent Type and Position Influence its Properties, Chem. – Eur. J. 21 (2015) 17728–17738. doi:10.1002/chem.201502127.

[35] H. Liu, S. Ryu, Z. Chen, M.L. Steigerwald, C. Nuckolls, L.E. Brus, Photochemical Reactivity of Graphene, J. Am. Chem. Soc. 131 (2009) 17099–17101. doi:10.1021/ja9043906.

[36] M.M. Lucchese, F. Stavale, E.H.M. Ferreira, C. Vilani, M.V.O. Moutinho, R.B. Capaz, C.A. Achete, A. Jorio, Quantifying ion-induced defects and Raman relaxation length in graphene, Carbon. 48 (2010) 1592–1597. doi:10.1016/j.carbon.2009.12.057.

[37] Z. Xia, F. Leonardi, M. Gobbi, Y. Liu, V. Bellani, A. Liscio, A. Kovtun, R. Li, X. Feng, E. Orgiu, P. Samorì, E. Treossi, V. Palermo, Electrochemical Functionalization of Graphene at the Nanoscale with Self-Assembling Diazonium Salts, ACS Nano. 10 (2016) 7125–7134. doi:10.1021/acsnano.6b03278.

[38] C.-J. Shih, Q.H. Wang, Z. Jin, G.L.C. Paulus, D. Blankschtein, P. Jarillo-Herrero, M.S. Strano, Disorder Imposed Limits of Mono- and Bilayer Graphene Electronic Modification Using Covalent Chemistry, Nano Lett. 13 (2013) 809–817. doi:10.1021/nl304632e.

[39] H. Hajova, Z. Komínková, A. Santidrian, O. Frank, L. Kubac, F. Josefik, M. Kalbac, Preparation and Charge-Transfer Study in a Single-Walled Carbon Nanotube Functionalized with Poly(3,4-ethylenedioxythiophene), J. Phys. Chem. C. 119 (2015) 21538–21546. doi:10.1021/acs.jpcc.5b06619.

[40] E. Poverenov, M. Li, A. Bitler, M. Bendikov, Major Effect of Electropolymerization Solvent on Morphology and Electrochromic Properties of PEDOT Films, Chem. Mater. 22 (2010) 4019–4025. doi:10.1021/cm100561d.

[41] T. Bashir, F. Bakare, B. Baghaei, A.K. Mehrjerdi, M. Skrifvars, Influence of different organic solvents and oxidants on insulating and film-forming properties of PEDOT polymer, Iran. Polym. J. 22 (2013) 599–611. doi:10.1007/s13726-013-0159-x.

[42] X. Crispin, S. Marciniak, W. Osikowicz, G. Zotti, A.W.D. van der Gon, F. Louwet, M. Fahlman, L. Groenendaal, F. De Schryver, W.R. Salaneck, Conductivity, morphology, interfacial chemistry, and stability of poly(3,4-ethylene dioxythiophene)–poly(styrene sulfonate): A photoelectron spectroscopy study, J. Polym. Sci. Part B Polym. Phys. 41 (2003) 2561–2583. doi:10.1002/polb.10659.

[43] L. Kavan, L. Dunsch, Spectroelectrochemistry of Carbon Nanostructures, ChemPhysChem. 8 (2007) 974–998. doi:10.1002/cphc.200700081.

[44] S. Garreau, G. Louarn, J.P. Buisson, G. Froyer, S. Lefrant, In Situ Spectroelectrochemical Raman Studies of Poly(3,4-ethylenedioxythiophene) (PEDT), Macromolecules. 32 (1999) 6807–6812. doi:10.1021/ma9905674.

[45] M. Kalbac, A. Reina-Cecco, H. Farhat, J. Kong, L. Kavan, M.S. Dresselhaus, The Influence of Strong Electron and Hole Doping on the Raman Intensity of Chemical Vapor-Deposition Graphene, ACS Nano. 4 (2010) 6055–6063. doi:10.1021/nn1010914.

[46] J. Yan, Y. Zhang, P. Kim, A. Pinczuk, Electric Field Effect Tuning of Electron-Phonon Coupling in Graphene, Phys. Rev. Lett. 98 (2007) 166802. doi:10.1103/PhysRevLett.98.166802.

[47] M. Lazzeri, F. Mauri, Nonadiabatic Kohn Anomaly in a Doped Graphene Monolayer, Phys. Rev. Lett. 97 (2006) 266407. doi:10.1103/PhysRevLett.97.266407.

[48] Z. Komínková, M. Kalbáč, Extreme electrochemical doping of a graphene–polyelectrolyte heterostructure, RSC Adv. 4 (2014) 11311–11316. doi:10.1039/C3RA44780D.

[49] Z. Komínková, M. Kalbáč, Raman spectroscopy of strongly doped CVD-graphene, Phys. Status Solidi B. 250 (2013) 2659–2661. doi:10.1002/pssb.201300070.

[50] C.-F. Chen, C.-H. Park, B.W. Boudouris, J. Horng, B. Geng, C. Girit, A. Zettl, M.F. Crommie, R.A. Segalman, S.G. Louie, F. Wang, Controlling inelastic light scattering quantum pathways in graphene, Nature. 471 (2011) 617–620. doi:10.1038/nature09866.

[51] K. Nagashio, T. Nishimura, K. Kita, A. Toriumi, Systematic Investigation of the Intrinsic Channel Properties and Contact Resistance of Monolayer and Multilayer Graphene Field-Effect Transistor, Jpn. J. Appl. Phys. 49 (2010) 051304. doi:10.1143/JJAP.49.051304.

[52] A. Elschner, ed., PEDOT: principles and applications of an intrinsically conductive polymer, CRC Press, Boca Raton, Fla., 2011.

**EDOT polymerization at photolithographically patterned functionalized graphene**

High-resolution PEDOT structures were fabricated by selective growth of the polymer on photolithographically patterned functionalized graphene. The composite exhibits uniform polymer thickness and distribution down to the micrometer scale. Efficient graphene-PEDOT electronic communication was demonstrated by several electrochemical methods. The material exhibit interesting features for fabrication of (opto-)electronic devices due to its electron-blocking/hole-transporting properties.

P. Kovaříček, K. Drogowska, Z. Komínková Melníková, V. Blechta, Z. Bastl, D. Gromadzki, M. Fridrichová, M. Kalbáč*

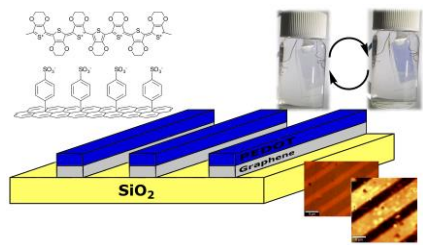